# Anisotropic nanoscale wrinkling in solid state substrates

*Maria Caterina Giordano and Francesco Buatier de Mongeot**

Dipartimento di Fisica - Università di Genova, Via Dodecaneso 33, I-16146 Genova, Italy.
*Corresponding author: buatier@fisica.unige.it

## Abstract

Pattern formation induced by wrinkling is a very common phenomenon exhibited in soft-matter substrates. In all these systems wrinkles develop in presence of compressively stressed thin films lying on compliant substrates. Here we demonstrate the controlled growth of self-organized nanopatterns exploiting a wrinkling instability on a solid-state substrate. Soda-lime glasses are modified in the surface layers by a defocused ion beam which triggers the formation of a compressively stressed surface layer deprived of alkali ions. When the substrate is heated up near its glass transition temperature, the wrinkling instability boosts the growth rate of the pattern by about two orders of magnitude. High aspect ratio anisotropic ripples bound by faceted ridges are thus formed which represent an optimal template for guiding the growth of large area arrays of functional nanostructures. We demonstrate the engineering over large $cm^2$ areas of quasi-1D arrays of Au nanostripe dimers endowed with tunable plasmonic response, strong optical dichroism and high electrical conductivity. These peculiar functionalities allow to exploit these large area substrates as active metamaterials in nanophotonics, biosensing and optoelectronics.

Pattern formation induced by mechanical wrinkling/buckling of surface layers and thin films is a very common phenomenon in nature, e.g. on dried fruit or human skin, or in polymers.[1-3] Moreover it is an important issue in many applications ranging from biology[4] to super-hydrophobic surfaces,[5] electronics[6,7] and photonics.[8,9] In all these systems wrinkles develop at the surface of compressively stressed thin films lying on top of compliant substrates.[10] Since the characteristic wavelength of the pattern depends on the thickness of the compressed film and on the ratio between elastic modulus of the substrate and of the film,[2] a wide range of periodicities can be achieved, spanning the entire range from nano- to macro-scale.[11]

In the case of polymeric materials different methods capable to induce stress at the surface have been developed, such as surface irradiation with ions or photons[12] and wet chemical treatments.[13] Under these conditions surface chemical oxidation occurs and crosslinks are formed in the surface layer thus inducing a compressive stress that can be vertically relaxed by wrinkling. In this way large area substrates can be patterned in a self-organized and maskless fashion but the control of the pattern morphology is limited[14] and undulations with relatively low height/width aspect ratio, in the range of 1/10, are generally obtained.[2,8] The competition of the elastic parameters of the skin and of the bulk support determine the wrinkles wavelength as described by the continuum model reported in the Methods section.[11,15]

In view of applications of relevance in fields ranging e.g. from photonics[16-18] to optoelectronics[19-21] and biotechnologies[22,23] it would be desirable to develop cost-effective methods which enable large area nanoscale wrinkling also at the surface of inorganic solid state materials. Unfortunately, the characteristic stiffness of this class of substrates generally hinders mass transport and the vertical relaxation of the top interface driven by the surface stress field. A notable exception is found when compressively stressed metal films are deposited on top of compliant polymer substrates which indeed exhibit a wrinkling instability.[7,24]

Alternatively, large area anisotropic periodic patterns can be induced on a broad class of solid state substrates ranging from dielectrics[25,26] and semiconductors[27,28] to metals[29,30] exploiting a self-organized technique based on defocused Ion Beam Sputtering (IBS).[31-33] Under this condition however, the kinetics of pattern formation is generally slow and induces the growth of shallow nanoripples in amorphous substrates, while faceted nanostructures oriented along the symmetry directions are induced only in the case of crystalline materials.[34,35]

In this work we describe a novel regime of self-organized nanopatterning driven by an ion-assisted wrinkling instability on glass substrates, a relevant material in optoelectronic applications and a prototype among solid-state amorphous substrates. Unexpectedly, the wrinkling instability develops only if high substrate temperatures are employed during ion irradiation, defying the prevailing assumption that considers smoothing of the nanostructures driven by thermally activated diffusion processes.[31-33] Highly ordered one-dimensional ripples with a height-to-width aspect ratio approaching unity develop with a vertical dynamic which is boosted by more than one order of magnitude with respect to low temperature experiments, while the timescale of pattern formation is reduced by one order of magnitude. Unexpectedly, in the late stages of the wrinkling process, the amorphous glass templates evolve into an asymmetric sawtooth profile with slope selected facets, thus opening a new scenario for self-organized nanopatterning of dielectric substrates.

The so formed self-organized nanopatterns indeed represent natural templates for e.g. driving the growth of quasi 1-dimensional (1D) gratings [36-40] endowed with peculiar photonic, optoelectronic and biosensing functionalities. A strong advantage in terms of scalability of the process towards novel applications and technologies is represented by the short nanofabrication time, in the range of minutes, and by the large functionalized area, in the range of cm$^2$.

Defocused ion beam sputtering of soda lime glass substrates at an off normal incidence angle $\theta=35°$ induces the formation of a partially ordered 1-dimensional ripple pattern as shown by the Atomic Force Microscopy (AFM) topography of **Figure 1(a)**. The nanopattern develops uniformly in a self-organized fashion over large cm$^2$ areas when ion bombardment is prolonged up to a sputtering time $t=3600$ s and the substrate temperature is kept in the range below 550 K,[25] where athermal processes dominate the relaxation dynamics. This observation is in agreement with the theoretical models[31] that predict the formation of a ripple pattern with a wave vector oriented parallel to the ion beam projection under the present irradiation conditions.

A numerical analysis of the AFM image shown in panel 1(a) provides an RMS roughness value $w$ of 3 nm and a typical ripple wavelength $\lambda$ of 140 nm, in agreement with the experimental state of the art of ion beam induced glass nanopatterns near room temperature.[25,36,39] Surprisingly, if the glass substrate is heated up above 600 K while keeping constant the ion fluence, a more than tenfold amplification of the vertical dynamic of the pattern occurs, as highlighted by the AFM topography in Figure 1(b) which represents a ripple pattern formed by IBS on a glass substrate heated up at 680 K. The monotonic amplification of the RMS surface roughness $w$ from 3 nm to 40 nm - Figure 1(c) - starts for T above 520 K (see AFM images and line profile at T=590 K and 620 K in Fig SI2) and continues up to 790 K, where a remarkable maximum of $w$ around 40 nm is observed. Corresponding, as shown in Figure 1(d), the average ripple wavelength $\lambda$ increases with temperature together with the typical ridge elongation, and the lateral coordination of the ripples is gradually enhanced (see 2D self-correlation patterns in Figure SI1), achieving the highest degree of order at 880 K.

These observations are at variance with prevailing thermodynamic arguments which predict that the enhanced diffusion at high T (either thermally activated or ion enhanced) should favour surface smoothing rather than roughening, and suggest the presence of a different destabilization mechanism. In order to address this issue we performed IBS experiments at $\theta=35°$ and T=680 K as a function of the ion fluence (Figure 2). The primitive shallow ripples already develop after an irradiation time as low as 121 s forming the pattern represented in Figure 2(a), characterized by $w$ =1 nm and $\lambda$= 75 nm. By slightly increasing the irradiation time up to 207 s - Figure 2(b) - a well defined ripple pattern develops with almost constant periodicity λ =70 nm and a strong increase of the vertical dynamic up to 25 nm ($w$ = 7.7 nm). After further increasing the ion dose beyond 250 s a different dynamic regime is accessed and the periodicity, length and lateral order of the ripples gradually increase as shown by the AFM topographies in Figure 2(c),(d) respectively corresponding to $t$ = 450 s and 1800 s.

To quantitatively evaluate the dynamic trend, in Figure 3(a),(b) we respectively plot the RMS roughness $w$ and the wavelength $\lambda$ of the ripple pattern versus the sputtering time $t$. The data relative to IBS experiments performed at 680 K (red squares) are compared with the experiment performed at low temperature (520 K, blue dot). It is worth noting that at high T, values of surface roughness comparable to the low T case are achieved with an ion fluence reduced by an order of magnitude, while at the highest fluency studied (t=3600 s) the roughness increases by an order of magnitude in comparison to the low T case [25]. As qualitatively suggested by the images of Figure 2(a),(b), when sputtering time $t$ increases from 65 s up to 207 s, $w$ exhibits a strong increase by more than one order of magnitude from 0.5 nm to 7.7 nm. The dynamics of the surface roughness follows the exponential trend $w(t)=w_0 \times e^{rt}$ with $w_0$=0.14 nm and growth rate $r$=1.8×10$^{-2}$ s$^{-1}$, corresponding to the orange line shown in Figure 3(a). During the same time interval the wavelength $\lambda$ remains constant in the range 70±20 nm - Figure 3(b). This behaviour is in qualitative agreement with the continuum model by Bradley Harper (BH) which describes the first stages of the IBS process when the small amplitude approximation can be applied (see Theoretical Model -Equation 2,3).[28,31-33]

By further increasing $t$ above 200 s, we observe that a power law trend sets in both for $w$ and $\lambda$ (Figure 3(a),(b)). Indeed, the small amplitude approximation is no longer valid in this regime and further non-linear mechanisms contribute to pattern formation. In particular, above $t$=200 s roughness and periodicity respectively increase as $w(t)=w_0 \times t^\alpha$ with $\alpha$=0.6±0.1 and as $\lambda(t)=\lambda_0 \times t^\gamma$ with $\gamma$=0.57±0.2. The power law scaling exponents $\alpha$ and $\gamma$ are comparable within error indicating that the shape and aspect ratio of the ripple cross section is preserved in the non-linear growth regime at high fluencies. Furthermore the exponents $\alpha$ and $\gamma$ are almost halved with respect to similar IBS experiments performed at low temperatures.[25] These observations confirm that the extraordinary vertical amplification observed at high T takes mainly place during the first stages of the IBS process, when the amplitude scales exponentially. Such a dynamic scaling suggests that the observed surface instability sets-in very rapidly and soon overtakes the conventional curvature dependent BH erosive instability. We stress however that the BH erosive instability plays an essential role in triggering surface destabilization and in determining the orientation of the ripple ridges with a wavevector parallel to the ion projection, due to the off-normal incidence conditions. [31]

The present scenario could be reconciled with the action of a wrinkling instability in presence of a compressively stressed surface layer, similarly to the case of wrinkled polymer films.[11] Indeed previous reports describe the formation of a compressively stressed surface layer following $Ar^+$ ion irradiation in soda-lime glasses.[41] The modification of the elastic properties of the compressed glass surface layers is mainly determined by the depletion of Na ions and by enrichment in Ca –Figure SI4-  from a near surface region which extends for a significant depth $h$ in the range of 15 nm. The latter figure goes well beyond the projected ion range, which reads about 3 nm for Ar ions at 800 eV, due to temperature enhanced migration of dopant ions under our high T experimental conditions. [42]

The simple continuum model, [11] which applies to wrinkled polymer films, allows to set an order of magnitude for the wavelength of the peculiar glass ripples assuming that the driving force is the compressive surface stress, induced by ion modification of the chemical composition. Using Equation 1 in Theoretical Model we derive a ripple wavelength $\lambda_c \cong$ 67 nm, considering a thickness of the compressively stressed sodium-depleted skin layer $h\sim15nm$ and the elastic parameters characteristic of soda lime glasses. The resulting value of $\lambda_c$ is in good qualitative agreement with the experimental value $\lambda$= 70±20 nm corresponding to the ripples periodicity in the early stages of ion irradiation, before the non-linear regime sets in for sputtering times $t$ exceeding 250 s - Figure 3(b).

We underline that the wrinkling instability can drive the formation of high amplitude glass patterns only if the compressively stressed top film is supported on a compliant foundation. In the present context, the high stiffness and viscosity of the amorphous glass foundation for temperatures below 500 K hinders mass transport and inhibits the vertical amplifications of the glass wrinkles (Figure 1(a)). Conversely, in the high temperature regime (700-800 K) the glass viscosity drops sharply by many orders of magnitude from $10^{19}$ to $10^{10}$ Pa×s,[43] thus providing an effective pathway for mass transport from the bulk towards the near surface layers.

The possibility of including an alternative destabilizing term (with positive sign) in Equation 3 - see Theoretical Model - in addition to curvature dependent roughening was also proposed.[44-46] The physical origin of this destabilizing term was attributed to anisotropic visco-elastic compressive deformation of the irradiated top layer and to the action of body forces following ion implantation down to a depth of few nm, comparable with the ion range. According to such models, the instability develops only for ion incidence angles exceeding 45° and the ripple wavelength follows a decreasing trend with increasing angles.[45,46] Such mechanism cannot be invoked to explain our observations since both occurrences above are not verified.

In our experiments we also observed that the wrinkling instability, so far described for soda-lime glass substrates, does not occur when an identical sputtering experiment is performed on a borosilicate glass substrate which is free from alkali ion dopants (Figure SI3). In the latter case one instead observes the formation of shallow parallel-mode ripples which are compatible with the conventional BH instability. All these observations thus confirm (i) that the wrinkling instability is related to the increase of the bulk modulus in the compressively stressed surface layer induced by a depletion of alkali ions and (ii) that the high aspect ratio pattern can develop only in presence of a compliant foundation which ensures sufficient mass transport from the bulk.

Interestingly, a careful inspection of Figure 2(d) reveals that ripples bound by slope selected facets are formed at the highest sputtering doses. Under this condition the local fluctuations of the ripples morphology induced by random arrival of single ions become averaged and an averaged ripple profile forms at the equilibrium. The faceting effect is highlighted by the AFM line profiles of Figure 3(c) which allow to directly evidence the evolution of the ripple profile from a rounded shape at $t = 207$ s (black profile) towards a sharply faceted morphology at increasing sputtering time $t = 450$ s (red profile) and $t = 1800$ s (blue profile).

In order to quantitatively study the slope selection process, in Figure 3(d) we plot the evolution with sputtering time of the statistical distribution of the local slopes, derived from the AFM topography, as a function of the sputtering time. At $t=120$ s the unimodal Gaussian distribution peaked at 0° with typical slopes below 5° corresponds to a very shallow height profile. By slightly increasing $t$ until 207 s (pink trace) a strong broadening of the distribution is observed representing slopes spread in the range -40° / +30°. By further increasing $t$ up to 225 s, i.e. beyond the linear regime, the slope distribution sharpens forming a characteristic peak around 25° whose intensity continuously increases as a function of the sputtering time (red curve at $t=450$ s, green curve at $t=1800$ s). This prominent feature is due to the preferential growth of slope selected facets tilted at 25° with respect to the horizontal plane. The opposite ripple ridges, directly facing the ion beam, instead form very steep slopes exceeding 50°. We stress that the slopes of these steep ridges can be underestimated due to tip convolution effects in such deep and steep grooves.

The unexpected growth of slope selected facets in an amorphous glass substrate defies a conventional interpretation. In the case of crystalline metal substrates faceting can in fact be explained by kinetic arguments in presence of Ehrlich-Schwoebel barriers[30,34] which inhibit adatom descent at step edges, while thermodynamically favoured crystal orientations can drive facet formation in crystalline semiconductors irradiated at high T above the recrystallization temperature.[35] Alternatively, faceting of amorphized Si was reported for low temperature sputtering at grazing angles due to shadowing and redeposition effects.[47] Under our non-grazing experimental conditions, however, both facets are illuminated by the ion beam and we can rule out the role of the latter mechanisms in the faceting process.

A recent theoretical work based on a continuum approximation [48] predicts the formation of sharply faceted morphologies during ion beam sputtering of amorphous substrates near normal incidence conditions, when parallel mode ripples are formed. The work predicts that a faceted asymmetric saw-tooth profile with the steepest ridge facing the ion beam should appear in the non-linear regime at high ion doses, analogously to our observations of Figure 3(c),(d). Under this framework the faceting process is explained in terms of a third order non-linear mechanisms which sets in at high doses, without the need of considering non-local redeposition and shadowing effects. Though the continuum model does not allow to directly extract a quantitative comparison with the present experimental data, nevertheless the qualitative picture presents some significant agreements which deserve further theoretical investigation.

The faceted glass substrates formed by anisotropic wrinkling represent a natural platform for the templated growth of multifunctional nanostructures over large areas. As an example, in Figure 4 we show nanostripe (NS) arrays endowed with tunable plasmonic response and electrical transport functionality, fabricated in a single-step maskless process. The confinement over $cm^2$ areas of Au nanostripe arrays proceeds via glancing-angle deposition ($\varphi=50°$, local thickness $t=30$ nm) as shown in the sketch view of Figure 4(a). Geometric shadowing of the steeper ripple ridges allows to selectively confine laterally separated Au NSs on the wider facets, which are tilted at the characteristic slope of about 25°. The top-view Scanning Electron Microscopy (SEM) image of Figure 4(b), acquired with the backscattered electron detector, evidences the sequence of Au stripes (bright areas supported on the wide ripple facets) alternated by uncovered glass portions (dark areas in correspondence to the steep narrow facets). The optical transmittance spectra shown in Figure 4(d), measured for incident light polarization either parallel (TE-red dashed line) or perpendicular (TM-blue line), reveal a broadband dichroic optical response extending from the Near-UV to the Near-IR. For TE polarization the NS array behaves as a semi-transparent Au film with the characteristic inter-band transition maximum peaked at about 530 nm wavelength. Conversely, for TM polarization a pronounced minimum is detected at about 570 nm due to excitation of a Localized Surface Plasmon (LSP) resonance along the transverse axis of the NSs.[38,49] The plasmonic response of the NS array can be easily tailored [50] either by changing the thickness of the nanostripes (Au dose) or by changing their width (i.e. by evaporating the Au atoms on the 'narrower' and steeper facets) as shown in Figure SI5. The quasi-1D NS arrays represent also a natural platform for the conformal growth of more complex 3-dimensional architectures. As an example, in the SEM cross sectional image of Figure 4(c), we show the vertical confinement of tilted Au-silica-Au nanodimer arrays (φ=50°, local thickness $t_{Au}$=30 nm- $t_{SiO2}$=44 nm- $t_{Au}$=30 nm). The Au stripes (brighter regions) are vertically disconnected by a thin silica layer (the dark stripes in the gap region). Under this condition the optical response of the plasmonic substrate results substantially modified by plasmon hybridization in the Au-silica-Au dimers. The TM-polarized transmission spectra for Au NS (blue line), silica capped Au-NS (grey line) and Au-silica-Au nanodimer arrays (green line) are compared in Figure 4(d). After encapsulating a single nanostripe with a silica layer a red-shift of the LSP resonance, from 570 nm to 600 nm, is detected due to an increase of the effective refractive index in the surroundings of the NS. Upon deposition of the second Au nanostripe, a strong change of the optical response of the array is observed. The LSP resonance corresponding to the single NS splits into two hybridized modes, a blue shifted one peaked at 550 nm (corresponding to the in phase oscillation of the electric-dipole modes) and a red-shifted one at about 680 nm (corresponding to the out-of-phase antibonding oscillation of the NS dipole modes). [50,51] The ability to shift the plasmon resonance of Au nanostructures towards the NIR spectral range represents a clear added value e.g. in view of quantitative plasmon enhanced spectroscopies such as Surface Enhanced Raman Scattering (SERS).[52,53] We also stress that the tilted geometry of the nanowires with respect to the substrate plane confers additional functionalities to these metasurfaces e.g in non linear second harmonic generation applications [39,54,55] and in directional Light Extinction and Emission.[56]

In view of opto-electronic applications we stress that NS arrays similar to the pattern of Figure 4(b) form a percolated network which ensures good electrical conductivity over macroscopic distances while being semi-transparent in the VIS-NIR spectral range. The sheet resistance measured in-situ during Au deposition at φ=-50° - inset of Figure 4(f) - on the steeper ripples ridges, follows a very fast percolative behavior as a function of the deposited metal dose $d$. In particular, we show the sheet resistance plot as a function of the Au thickness $d$ that would be deposited on a flat surface under the same experimental conditions. The electrical contact is established for $d$>17 nm when the sheet resistance drops reaching values below 10 Ohm/sq. Under this condition the arrays are semi-transparent and support dichroic excitation of a localized plasmon resonance, in analogy to the spectra of Figure 4(d). The electrical transport of these self-organized arrays is thus competitive with respect to the best transparent conductive oxides based on ITO with the additional advantage that the nanoelectrodes are endowed with and intrinsic plasmonic functionality.

In conclusion we have demonstrated the possibility to induce large area faceted nanopatterns on a solid state glass substrate by exploiting a wrinkling instability which presents strong analogies with soft-matter systems. The self-organized nanofabrication method employs a defocused ion beam to trigger the formation of a compressively stressed surface layer deprived of alkali ions. A stress induced wrinkling instability occurs only when the substrate is heated up near its glass transition temperature, where viscosity drops sharply by several orders of magnitude. Under these conditions the compliancy of the substrate allows the vertical relaxation of the surface stress by wrinkling. Concurrently the uniaxial orientation of the ripples is determined by the erosive ion beam sputtering instability, that also drives the growth of highly ordered facets as long as several microns. The wrinkled glass templates allow to engineer the assembly of quasi-1D metal nanostripe arrays with tailored plasmonic and electrical response. Additionally, metallic dimer arrays supporting hybridized plasmonic modes can be obtained by exploiting the same platform. This represents a crucial step towards the development of large area substrates based on metamaterials whose applications span from nanophotonics[57] and non-linear optics,[58,59] to biosensing, [60,61] and flat optics.[62]

## Experimental Section

*Experimental Setup*

Soda-lime microscope glass slides are nanopatterned over large $cm^2$ areas by defocused Ion Beam Sputtering (IBS). The ion bombardment is performed in a custom made Ultra High Vacuum setup equipped with a gridded multiaperture ion source (TECTRA) which generate a defocused $Ar^+$ ion beam (5N purity) at the energy E = 800 eV, the extraction grid is polarized at the voltage V= -200 V. In this condition the ion flux reaching the sample at normal incidence is about $9.4 \times 10^{15}$ ions $cm^{-2}$ $sec^{-1}$ and the beam is characterized by an angular divergence of about 5°. We perform IBS experiments at selected ion beam incidence angle θ=35° with respect to the surface normal direction. A biased tungsten filament ($V_{bias}$= -13 V), providing electrons by thermoionic emission, is placed close to the extraction grid in order to compensate surface charging due to ion implantation.

The substrate temperature during IBS is monitored with a K-type thermocouple fixed to the sample holder which is thermally anchored to the substrate by silver paste layer on the back side of the sample.

If the IBS is performed without previous annealing of the substrate, we observed ion induced heating during the first stages of the process with an increase of the substrate temperature from room temperature to about 520 K.

Alternatively, the glass substrate can be heated up during the IBS process in a controlled way recurring to a Firerod heater, placed in thermal contact with the manipulator. In this case the ion beam sputtering induces a smaller effect with a further temperature increase of the substrate of the order of 10%-15% during the early stages of the ion bombardment.

**Theoretical Model**

*Surface wrinkling of compressively stressed elastic film*

The continuum model developed for a compressively stressed film supported on compliant elastic foundation[11,15] allows to estimate the wavelength of the glass wrinkles according to the relation

$$\lambda_c = 2\pi h \left[ \frac{(1-v_f^2)E_s}{(1-v_s^2)E_f} \right]^{1/3} \qquad (1)$$

where $h$ is the thickness of the compressively stressed skin layer, $E_s$ and $E_f$ are the elastic moduli of the skin and of the bulk support respectively, while $v_f$ and $v_s$ are their Poisson ratios.

In the case of soda lime glass we calculated $\lambda_c \cong 60$ nm assuming in Equation 1 a skin layer of thickness $h\sim15nm$,[42] an elastic modulus of skin/foundation of $E_s \cong 7.40\times10^{10}$ Pa and $E_f \cong 6.84\times10^{10}$ Pa and a Poisson ratio $v_f \cong 0.2081$ and $v_s \cong 0.2196$. We stress that the $\lambda_c$ value must be considered just as an order of magnitude estimate, based on the variation in the concentration of the main glass dopants above 1% atomic concentration (Na, Mg and Ca) from Energy Dispersive X-ray spectroscopy (EDS) for pristine and ion irradiated glasses.

*Ion beam induced nanopatterning*

The prevailing models of pattern formation induced by sputtering are based on Bradley Harper (BH) theory[31] and its developments, which describe pattern formation in terms of the competition between ion induced erosion and thermally activated diffusive relaxation. By considering the ion beam projection on the sample plane parallel to the $x$ axis, the evolution of the vertical profile $h(x,t)$ as a function of time $t$, is described by the following differential equation:

$$\frac{\partial h}{\partial t} = v_0 - \frac{\partial v_0}{\partial \vartheta}\frac{\partial h}{\partial x} + S\frac{\partial^2 h}{\partial x^2} - B\nabla^4 h \qquad (2)$$

The first and the second term describe the average (local) erosion rate which depends on the average (local) incidence angle. The third term of the equation describes ion induced roughening dependent on the curvature of the surface profile, while the fourth order term $-B\nabla^4 h$ describe thermal activated diffusion and smoothing. This description can be applied at the early stages of the IBS process, i.e. small amplitude approximation, and end up with an exponential growth of nanoripples height as $h(x,t)=h_0\times e^{r(q)\times t}$.

Experiments on amorphous silica[63] and glass[64] substrates showed that also ion induced viscous flow – either confined to the surface or involving also the bulk - can contribute strongly to surface smoothing by including two further linear relaxation terms proportional respectively to the third and first spatial derivative respectively. Under these conditions and within the linear regime valid in the early stages of sputtering, the exponential growth rate $r(q)$ of harmonic modulations of the surface profile is thus described by the following formula:

$$r(q) = Sq^2 - Fq - F_S d^3 q^4 - Bq^4 \qquad (3)$$

Where $Sq^2$ is the ion induced roughening term, $-Fq - F_S d^3 q^4$ are respectively the ion induced bulk and surface viscous relaxation terms, while $-Bq^4$ is the temperature induce relaxation term. According to Equation 3 one would thus expect a decrease of the ripple growth rate $r(q)$ if the mobility of the surface defects is dominated by thermally or ion activated processes which contribute with a negative sign.

## Acknowledgements

Financial support is gratefully acknowledged from Ministero dell'Università e della Ricerca Scientifica (MIUR) through the PRIN 2015 Grant No. 2015WTW7J3, from Compagnia di San Paolo in the framework of Project ID ROL 9361 and from MAECI in the framework of the Italy-Egypt bilateral protocol.

# Figures

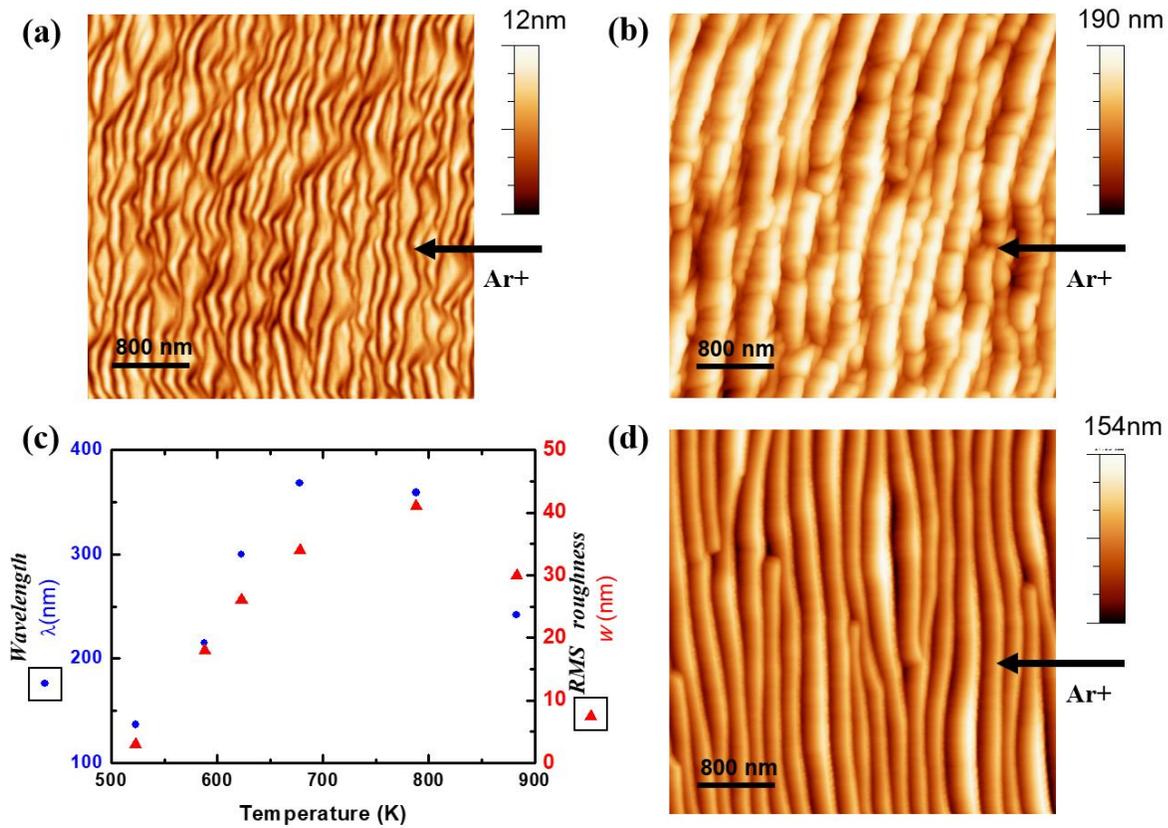

**Figure 1.** (**a, b, d**) AFM topographies of ripples patterns induced on soda lime glasses after 3600s of IBS at θ=35°. The substrate temperatures, T, respectively reads 520 K, 680 K, 880 K. (**c**) Wavelength λ and RMS roughness w of the ripple patterns plotted as a function of the substrate temperature.

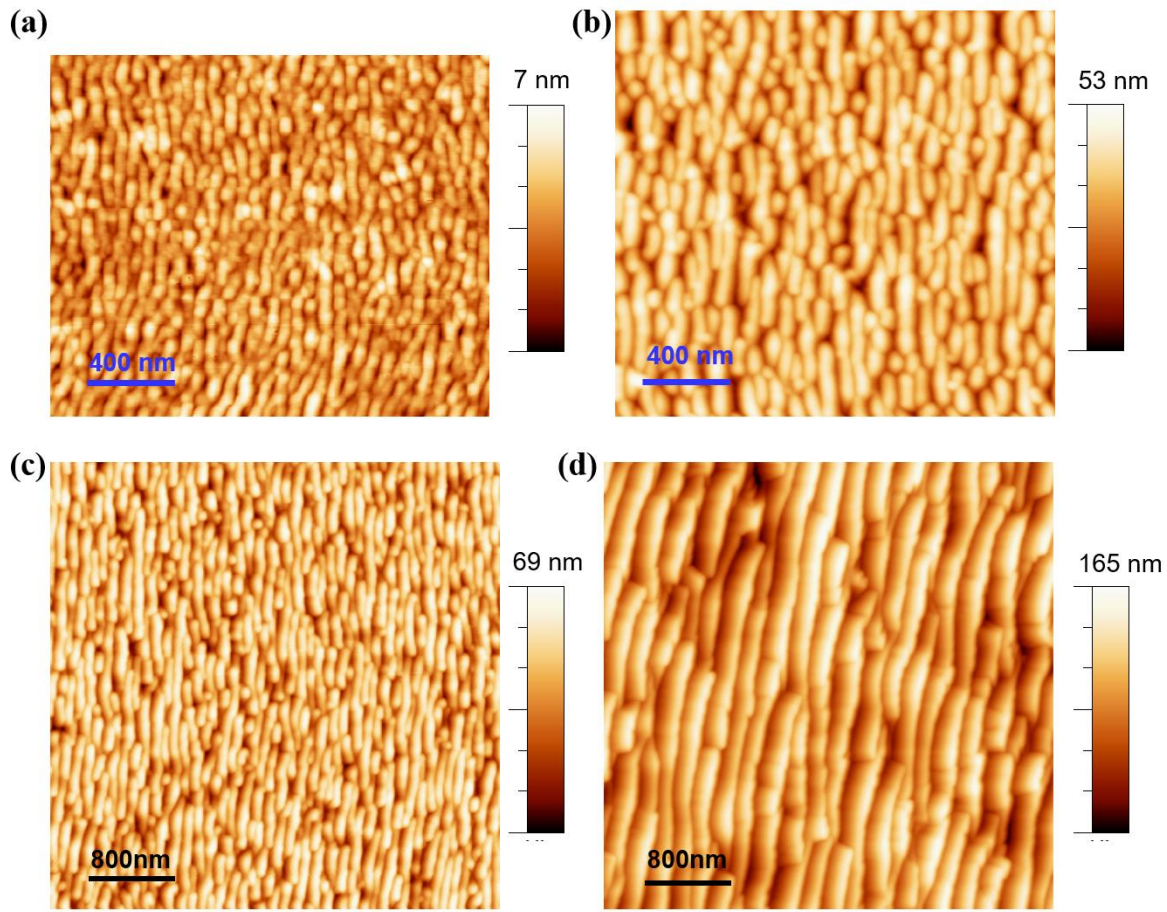

**Figure 2** AFM topographies of ripples patterns induced on soda lime glasses by IBS at θ=35° and T=680 K. The substrates have been irradiated at increasing ion fluencies which correspond to a sputtering time of (**a**) 121 s (9.3×10$^{17}$ ions cm$^{-2}$), (**b**) 207 s (1.59×10$^{18}$ ions cm$^{-2}$), (**c**) 450 s (3.5×10$^{18}$ ions cm$^{-2}$), (**d**), 1800 s (1.4×10$^{19}$ ions cm$^{-2}$) . In (c) and (d) the scan size is doubled.

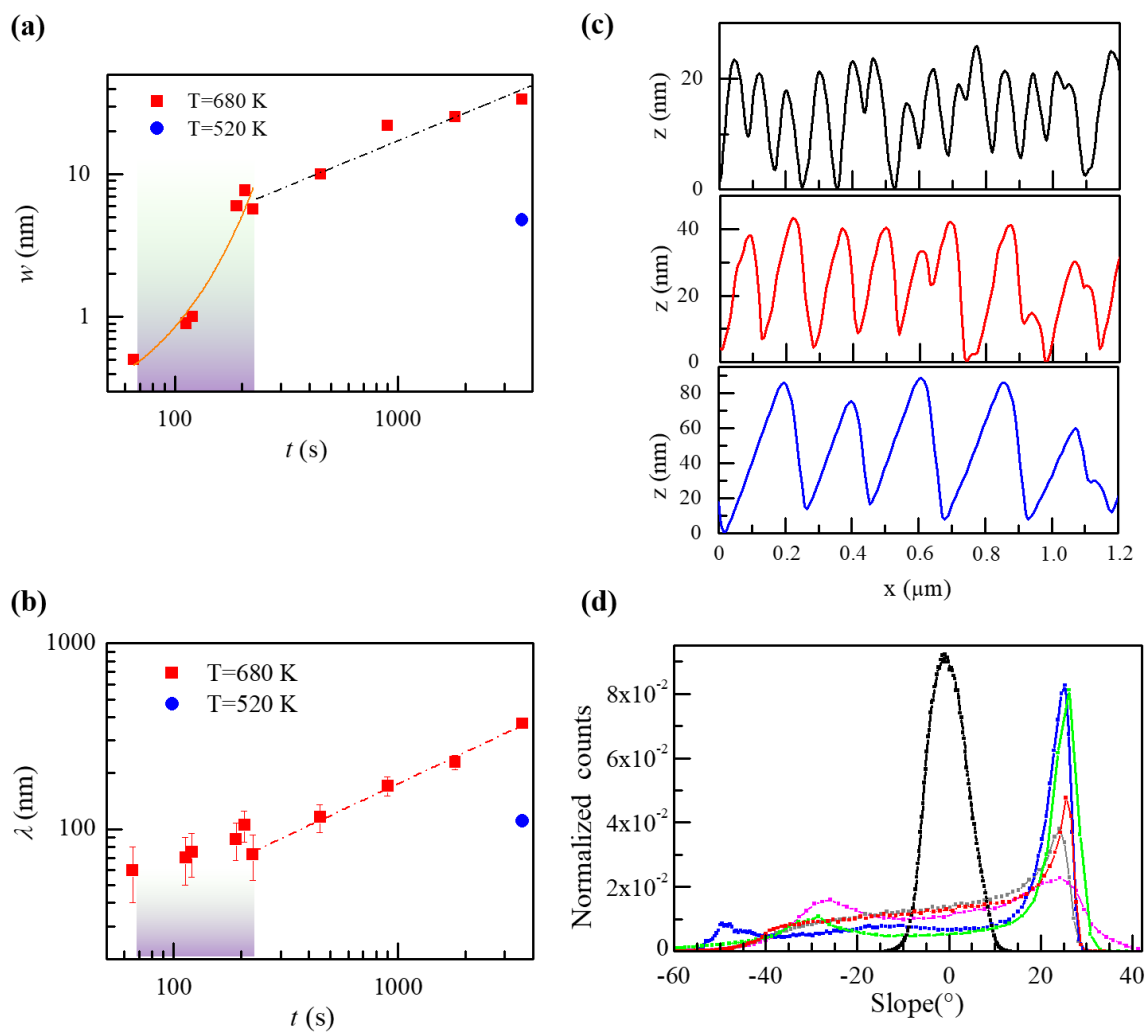

**Figure 3 (a, b)** The data of RMS roughness $w$ and the periodicity $\lambda$ of the ripple patterns are plotted as a function of the sputtering time $t$. The red and blue points respectively correspond to IBS experiments at 680 K and 520 K. **(c)** Selected AFM line profiles after sputtering at T=689 K for increasing times $t$=207 s (black line), $t$=450 s (red line), $t$=1800 s (blue line). **(d)** Histogram of slopes, evaluated with respect to the average plane, extracted by the AFM topography corresponding to increasing sputtering time $t$ at fixed temperature T=680 K and ion incidence angle $\theta$=35°. Black line ($t$=121 s), pink line ($t$=207 s), grey line ($t$=225 s), red line ($t$=450 s), blue line ($t$=1800 s), green line ($t$=3600 s).

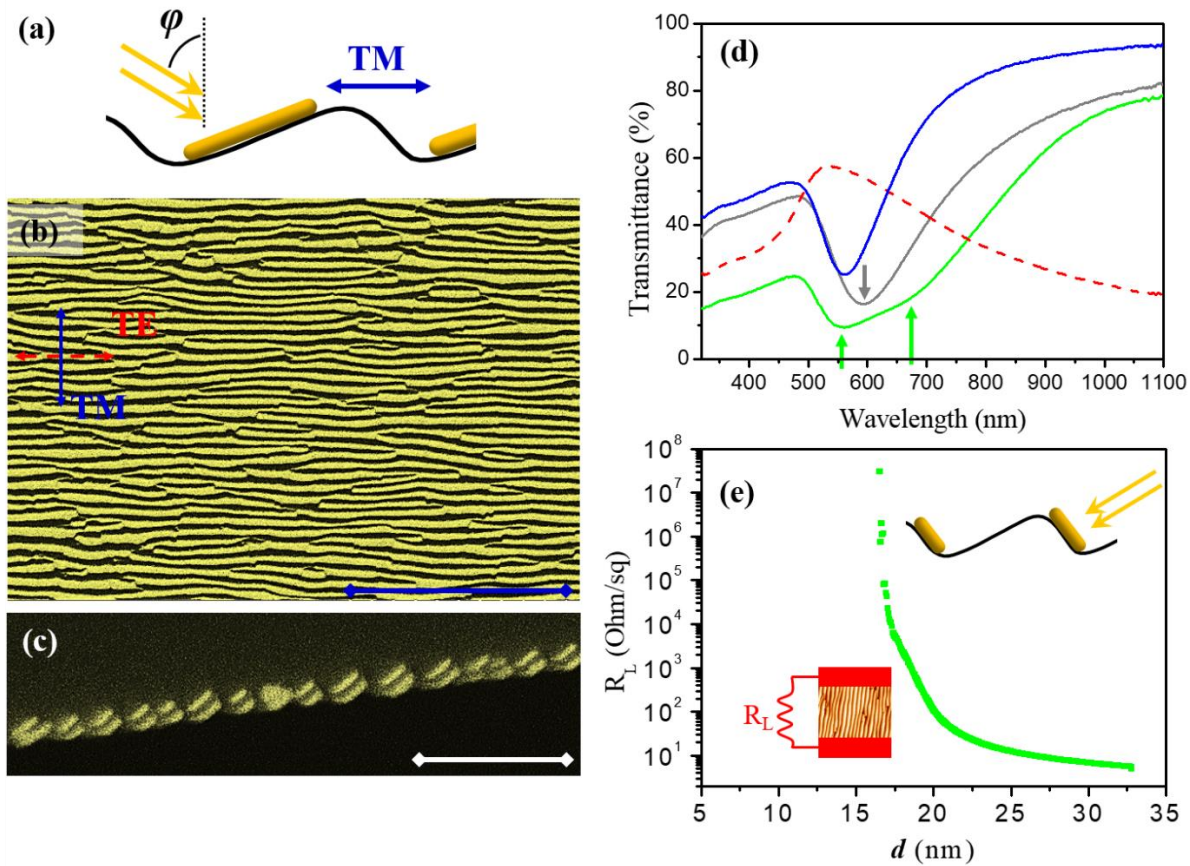

**Figure 4 (a)** Schematic view of the cross-section of the Au nanostripes (NS) confined on the faceted glass pattern by glancing metal deposition **(b,c)** SEM image of the Au NS arrays (top-view) and of the Au-silica-Au nanodimers (cross-section view), respectively. The blue and white scale bar correspond to 3 μm and 500 nm respectively. **(d)** Optical transmission spectra of the NS arrays detected under incident light polarization either parallel (TE-red dashed line) or perpendicular (TM-blue line) to the NS long axis. Also shown are the Optical spectra (TM–polarized) of the silica capped Au NS arrays (grey line) and of the Au-silica-Au dimer arrays (green line) highlighting the plasmonic hybridization in the case of the dimers. **(e)** Longitudinal sheet resistance of the NS arrays measured over a large area circuit (5×5 mm$^2$) during Au deposition at φ=-50° on the steep ripples ridges tilted at about 50° as a function of the thickness *d* of the Au film.

# Anisotropic nanoscale wrinkling in solid state substrates

*Maria Caterina Giordano and Francesco Buatier de Mongeot\**

Dipartimento di Fisica - Università di Genova, Via Dodecaneso 33, I-16146 Genova, Italy.

*Corresponding author: buatier@fisica.unige.it

The evolution of the surface topography in soda lime glasses as a function of the substrate temperature is highlighted by AFM images of Figure SI1 (a),(c),(e) and by the corresponding 2D self-correlation pattern shown in Figure SI1 (b),(d),(f). The glass templates are all induced under the same IBS conditions, i.e. $Ar^+$ irradiation at $\theta=35°$ with an ion fluencies of $2.8 \times 10^{19}$ ions $cm^{-2}$, while the substrate temperature during the IBS process respectively reads 520 K, 680K and 880K as in Figure 1 (a),(b),(d) of the manuscript. A strong enhancement of the vertical dynamic from few nanometers up to 100 nanometers occurs with a gradual self-ordering of the ripple pattern. The intensity and the elongation of the central peak of the 2D self-correlation pattern gradually increases with temperature as well as the lateral peaks that are maximized at the highest temperature of 880 K. The enhanced surface diffusion induced at such high temperatures strongly promotes lateral coordination of nanoripples during growth while the ion induced wrinkling instability drives the ripples growth with enhanced vertical dynamics.

When the substrate temperature during the IBS reads between 590 K and 620 K an intermediate situation is observed, as highlighted by the AFM images of Figure SI2 (a),(c) corresponding to T=590K and 620K, respectively. The images and the corresponding ripples cross-section (Figure SI2(b),(d)) show a gradual increase of the vertical dynamic of the ripples with respect to the case of IBS at T=520 K (Figure SI1a). Indeed, the surface and bulk mobility start increasing within this temperature range allowing only a partial relaxation of the surface stress by wrinkling instability. The full development of the wrinkling mechanism can be observed only at higher surface and bulk mobility corresponding to sutface temperature of about 680 K (Figure SI1c).

The wrinkling instability and the facet growth observed in sputtered soda lime glass is a peculiar behavior of this material in the range of temperature here explored. In particular we observed that a borosilicate glass - Figure SI3 (a),(b) and a soda-lime glass (Figure SI3 (c),(d)) ion beam irradiated at $\theta=35°$ and at the substrates temperature of 680 K in the same experimental run exhibit a very different surface pattern. Both the AFM topography (Figure SI3 (a)) and the AFM line profile (Figure SI3 (b)) of the borosilicate glass highlight a reduced vertical dynamic in the range of 6-8 nm with a surface RMS roughness of the pattern of 1.5 nm. On the other hand, ripples nanostructure with vertical dynamics exceeding 100 nm and surface RMS roughness of 30 nm develop at the surface of the soda lime glass as shown in the AFM topography (Figure SI3(c)) and in the AFM line profile (Figure SI3 (d)). The strong differences observed in the surface pattern is attributed to different glass composition and in particular to the selective presence of alkali ions only in soda lime glasses. Only the latter substrates under IBS develop a compressively stressed surface layer deprived of alkali ions that drives the wrinkling instability and the growth of high aspect ratio nanoripples. In absence of the driving force responsible for the wrinkling instability in borosilicate glasses, one instead observes the formation of shallow parallel-mode ripples which are compatible with the conventional BH instability.

The Energy Dispersive X-ray (EDX) analysis performed on sputtered soda lime glasses (Figure SI4) highlights the effect of ion irradiation on the surface layer composition. The SEM image (Figure SI4 (a)) and the corresponding EDX map for Na concentration (Figure SI4 (a)) have been acquired on a portion of the glass surface partially screened from ions (red square) during the IBS process. The SEM image shows the ion erosion step at the edge of the screen while the EDX map highlights a reduction of the Na concentration on the ion irradiated surface with respect to the screened portion. An ion induced relative depletion of Na by 18% can be estimated by considering the EDX counts collected over screened and unscreened areas (green and blue areas respectively). Conversely an ion induced increase of Ca, and Mg relative concentration by about 33% and 4,7% is respectively detected after ion irradiation. Under these conditions an increase of the Young modulus is expected to take place in the irradiated surface layer that induces compressive stress accumulations at the surface able to drive the wrinkling instability. We stress that due to the large sampling range of the EDX measurement in the micrometer range, the composition of the deeper regions unaffected by the ion beam is averaged with that of the surface modified layers, thus resulting in an underestimation of the alkali depletion.

*Templates for confinement of quasi-1D plasmonic nanowire arrays*

The self-organized nanopatterned glasses represents optimal templates for confinement of highly anisotropic metallic gratings over large cm$^2$ area. In particular, laterally disconnected gold nanowire arrays can be prepared by glancing angle thermal deposition on the pre-patterned glass templates (Figure

SI5 (a)). The faceted ripples ridges enable confining highly ordered quasi-1D arrays of plasmonic nanostripes (NS) with selected tilt and width. Glancing angle Au deposition is performed perpendicularly to the ripple ridges and the Au beam is tilted at an angle φ with respect to the surface normal direction. Under these conditions the width of the Au NS cross section is imposed by the ripple template, while the local thickness can be easily tailored with the selected metal dose. If the metal deposition is performed at φ= 50° on the ripples facets tilted at 30° quasi- 1D NS arrays are obtained (SEM image of Figure SI5(b)), which are characterized by width of about 100 nm, imposed by the rippled template. The optical response of this substrate (Figure SI5(c)) is strongly anisotropic in the whole Near UV – Visible – Near IR spectrum (Extinction spectra of Figure SI5(c)). For incident light polarization parallel to the ripple ridges (TE-pol, green line-see also sketch in Figure SI5(b)) a drop of the transmittance below 530 nm wavelength is detected due to excitation of Au interband transitions, similarly to the optical behavior of a continuous thin Au film. For incident light polarization perpendicular to the ripple ridges (TM-pol, purple line) a pronounced transmission minimum is detected at 610 nm wavelength due to the excitation of a Localized Surface Plasmon (LSP) resonance along the cross section of the laterally disconnected Au NS. The plasmonic excitation induces a strong optical dichroism in the red and Near-IR spectrum where the quasi-1D gratings behave as wire grid polarizer.

Furthermore, the plasmonic response can be effectively tailored by depositing Au under a different incident angle φ= -70° on the opposite narrow facets tilted at about -50° (see histogram of Figure 3(d)). The SEM image and the corresponding optical response of the so formed Au NS arrays are shown in Figure SI5(d)-(e) respectively. A blue-shift of the LSP excitation from 610 nm wavelength (Figure SI5(c)) to 530 nm (Figure SI5(d)) is observed while a similar optical dichroism characterized the optical spectra. This effect is attributed to the strong change of the NS cross section. Indeed, while the Au thickness deposited on the ripples ridges is kept constant at about 20 nm, the width of the NS changes from about 100 nm (Figure SI5(b),(c)) to about 60 nm (Figure SI5(d),(e)) due to the strongly asymmetric ripple profile.

The self-organized nanopatterned glasses can thus be exploited as versatile templates for the large area confinement of quasi-1D metallic gratings endowed with high optical dichroism and tunable plasmonic functionalities in the VIS and Near-IR spectrum.

**SI Figures**

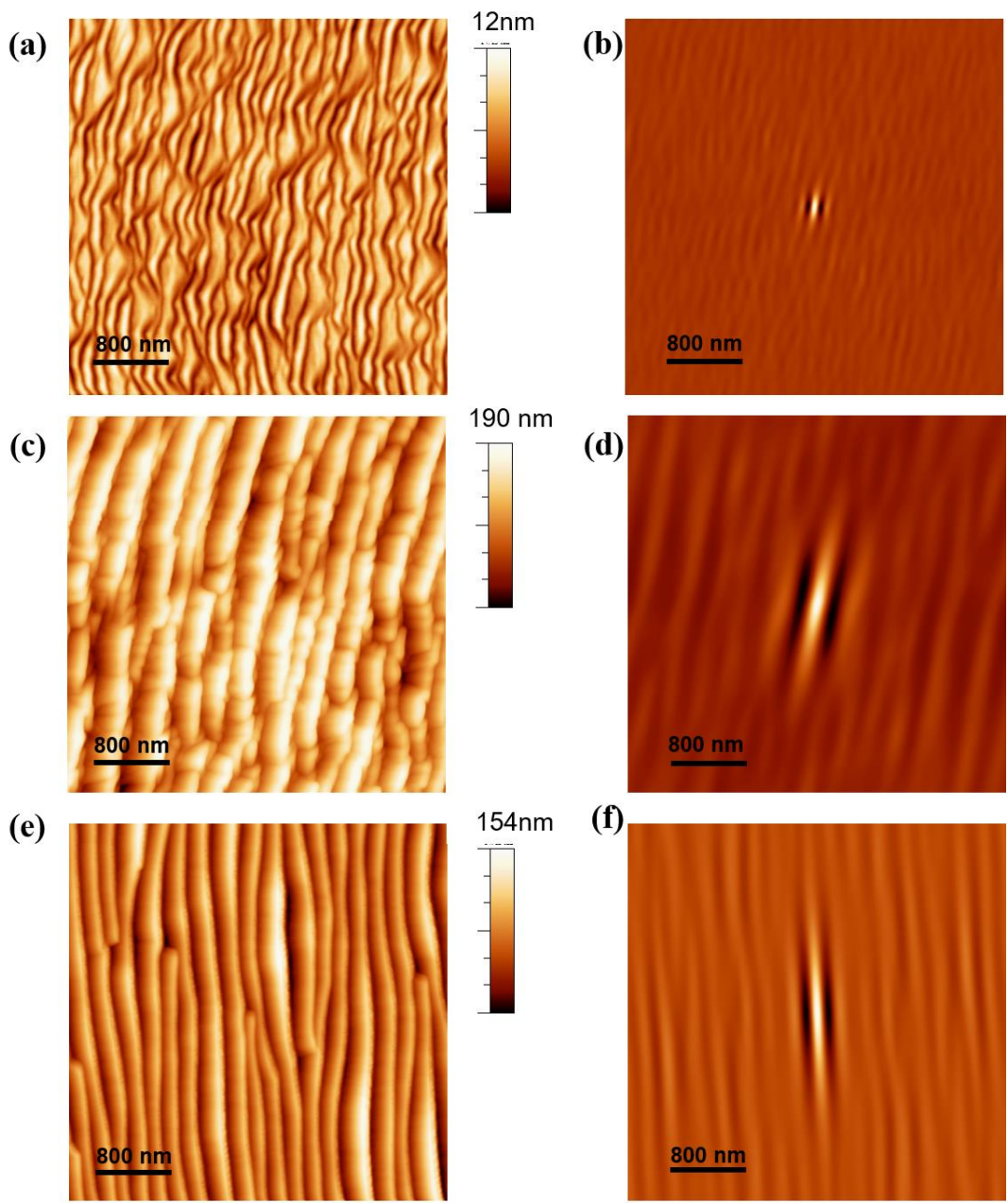

**Figure SI1 (a),(c),(e)** AFM topographies and **(b),(d),(f)** two dimensional self-correlation patterns of ripples patterns induced on sputtered soda lime glasses for a fixed sputtering time *t*=3600 s (ion fluence=2.8 ×10$^{19}$ ions cm$^{-2}$) at the substrate temperature of 520 K, 680K and 880K (shown in Figure 1 of the manuscript).

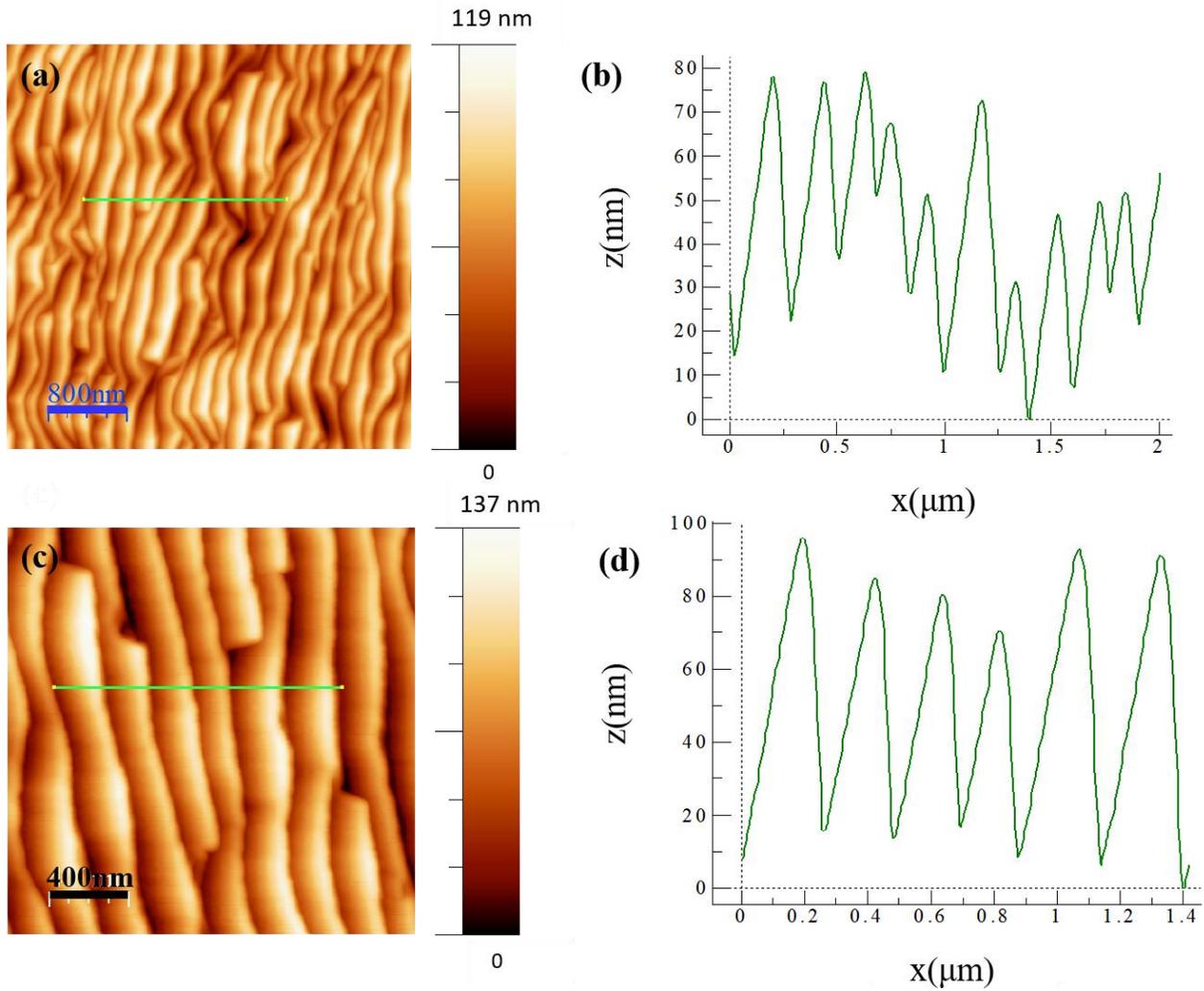

**Figure SI2 (a),(c)** AFM topographies and **(b),(d)** line profiles of ripples patterns obtained by IBS of soda lime glasses at the substrate temperature of 590 K and 620 K, respectively. The IBS was prolonged sputtering time $t$=3600 s (ion fluence=2.8 ×10$^{19}$ ions cm$^{-2}$) in both the cases.

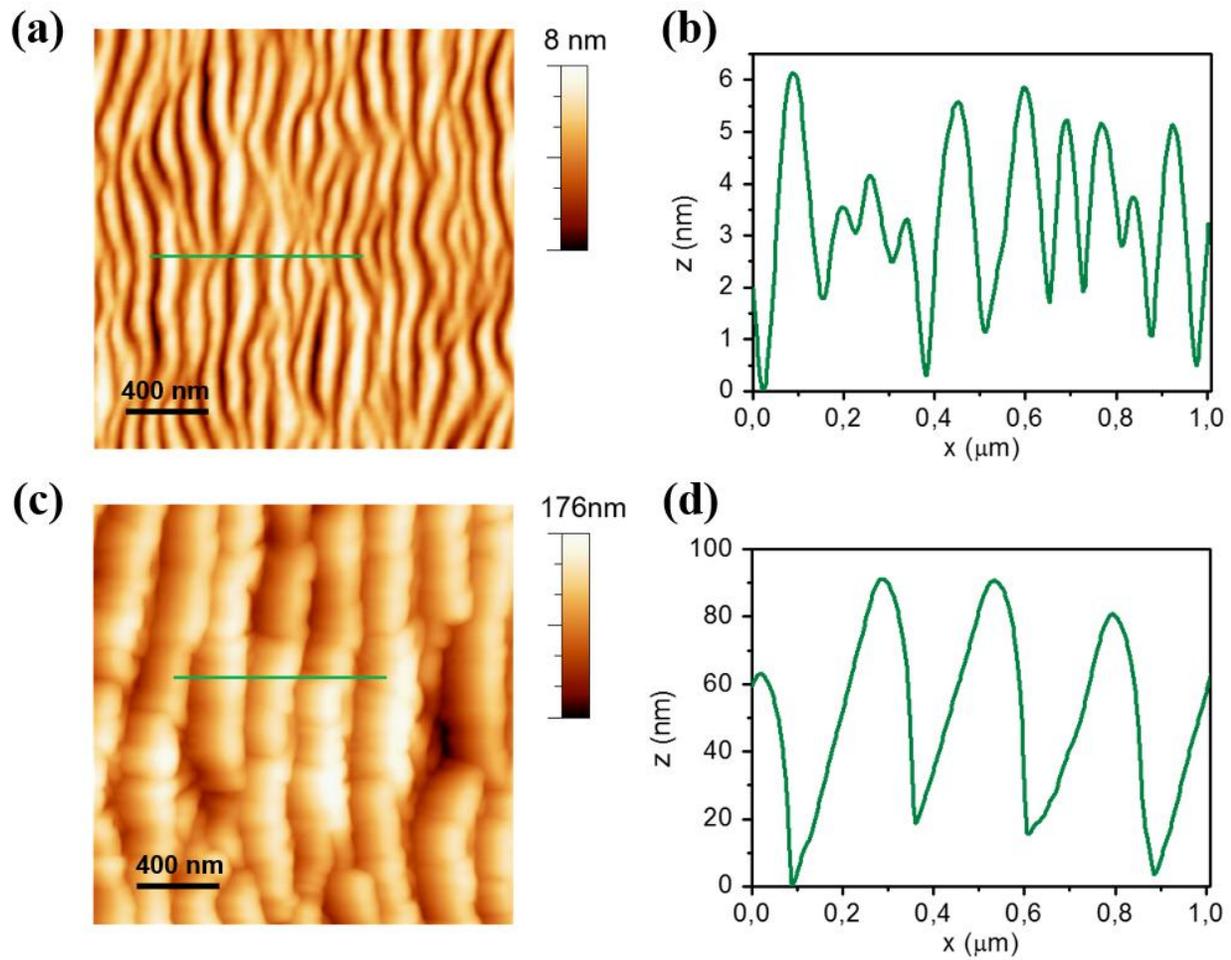

**Figure SI3** Rippled glass templates prepared by IBS at θ=35° and at the substrate temperature of 680K in the same experimental run on a borosilicate glass **(a),(b)** and on a soda lime glass **(c,d).** The sputtering time was prolonged up to $t$=3600 s (ion fluence=2.8 ×$10^{19}$ ions cm$^{-2}$).

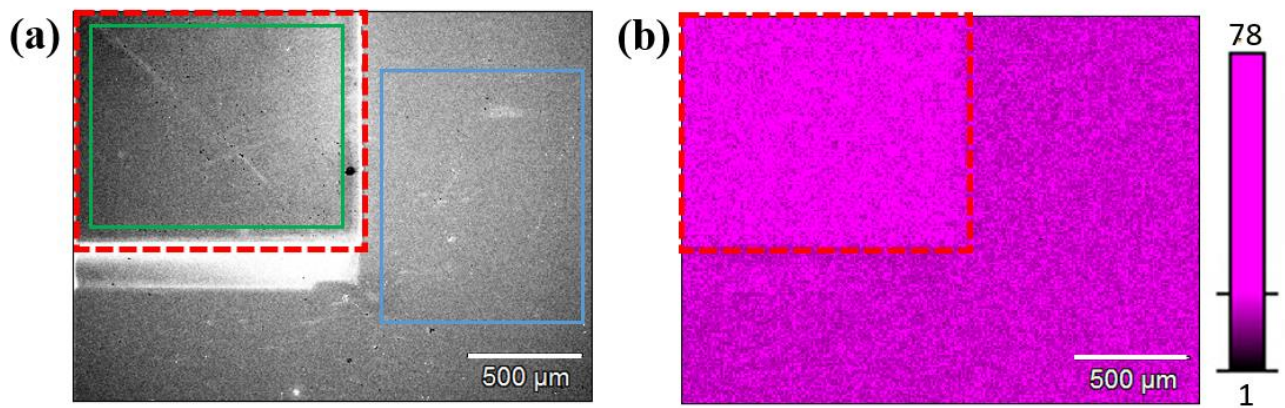

**Figure SI4 (a),(b)** SEM image and corresponding EDX map for Na concentration of a sputtered soda lime glass substrate where the portion of the surface delimited by the red line has been screened from ion irradiation . The green and blue rectangles represent the areas on the areas where the EDX counts has been integrated respectively on the ion-screened and sputtered glass region.

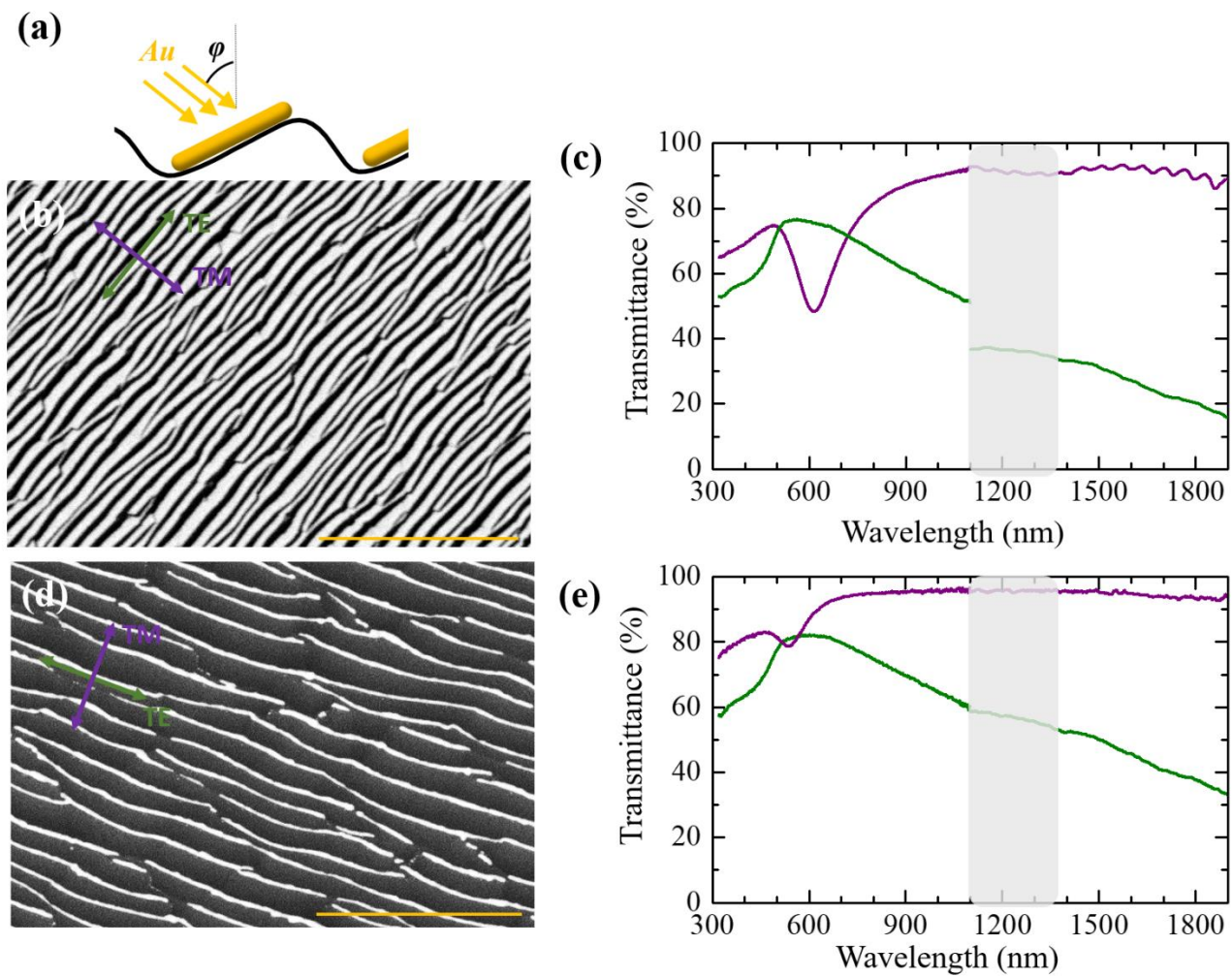

**Figure SI5 (a)** In-scale sketch of the glancing angle Au deposition procedure on the ripple glass template. **(b,c)** SEM image (scale bar corresponds to 3 μm) and optical extinction spectra of Au NSW arrays confined at φ=50° on the rippled template. **(d,e)** SEM image (scale bar corresponds to 2 μm) and optical extinction spectra of Au NS arrays confined at φ=-70° on the rippled template. The optical extinction has been measured for longitudinal polarization (TE-pol, green lines) and transversal polarization (TM-pol, purple lines).